\documentstyle[sprocl]{article}

\input{psfig}

\newcommand{\vk}{\mbox {\boldmath $k$\unboldmath}}
\newcommand{\vq}{\mbox {\boldmath $q$\unboldmath}}
\def\nn{\nonumber}

\def\Journal#1#2#3#4{{#1} {\bf #2}, #3 (#4)}

\def\AP{\em Ann. Phys.}
\def\EPJA{{\em Eur. Phys. J.} A}
\def\EPJC{{\em Eur. Phys. J.} A}
\def\NPA{{\em Nucl. Phys.} A}
\def\NPB{{\em Nucl. Phys.} B}
\def\PLB{{\em Phys. Lett.}  B}
\def\PPNP{\em Prog. Part. Nucl. Phys.}
\def\PR{\em Phys. Rep.}
\def\PRL{\em Phys. Rev. Lett.}
\def\PRC{{\em Phys. Rev.} C}
\def\PRD{{\em Phys. Rev.} D}
\def\PS{\em Phys. Scripta}
\def\RMP{\em Rev. Mod. Phys.}

\begin{document}

\title{SEARCH FOR NEW BARYON RESONANCES} 

\author{Bijan Saghai}

\address{Service de Physique Nucl\' eaire, DAPNIA - CEA/Saclay,\\
F-91191 Gif-sur-Yvette Cedex, France\\
E-mail: bsaghai@cea.fr
}

\author{Zhenping Li}

\address{Physics Department, Peking University,
Beijing 100871, P.R. China}

%%%%%%%%%%%%%%%%%%%%%%%%%%%%%%%%%%%%%%%%%%%%%%%%%%%%%%%%%%%%%%

\maketitle\abstracts{ Within a chiral constituent quark formalism,
allowing the inclusion of all known resonances, 
a comprehensive study of the recent $\eta$ photoproduction data on 
the proton up to $E_\gamma ^{lab} \approx$2 GeV is performed. This
study shows evidence for a new $S_{11}$ resonances and indicates the
presence of an additional missing $P_{13}$ resonance.  
}

%%%%%%%%%%%%%%%%%%%%%%%% Introduction %%%%%%%%%%%%%%%%%%%%%%%%
%
\section{Introduction}
%%%

For several decades, the baryon resonances have been 
investigated~\cite{PDG} 
mainly through partial wave analysis of the ``pionic'' processes 
$\pi N \to \pi N ,~\eta N$,~$\gamma N \to \pi N$, and to less
extent, from two pion final states.

Recent advent of high quality electromagnetic beams and 
sophisticated detectors, has boosted intensive experimental
and theoretical study of mesons photo- and electro-production.
One of the exciting topics is the search for new baryon resonances 
which do not couple or couple too weakly to the $\pi N$ channel.  
Several such resonances have been  predicted~\cite{Review,Cap92,GSV} 
by different QCD-inspired approaches, offering strong test of the underlying concepts.

Investigation of $\eta$-meson production {\it via} electromagnetic probes
offers access to fundamental information in hadrons spectroscopy. 
The properties of the decay of baryon resonances into $\gamma N$ 
and/or $N^* \to \eta N$ are intimately related to their internal 
structure~\cite{Cap92,IK,BIL}.
Extensive recent experimental efforts on $\eta$ 
photoproduction~\cite{Mainz,elsa,graal,Graal2000,CLAS2001} 
are opening a new era in this topic. 

The focus in this manuscript is to study all the recent 
$\gamma p \rightarrow \eta p$ data for $E_{\gamma}^{lab} <$ 2 GeV 
($W \equiv E_{total}^{cm} <$ 2.2 GeV) within a chiral constituent 
quark formalism based on the $SU(6)\otimes O(3)$ symmetry. 
The advantage of the quark model for meson photoproduction is 
the ability to relate the photoproduction data directly to the internal 
structure of the baryon resonances. Moreover, this approach allows
the inclusion of all of the known baryon resonances.
To go beyond the exact $SU(6)\otimes O(3)$ symmetry, 
we introduce~\cite{LS-1,LS-2}
symmetry breaking factors and relate them to the configuration mixing
angles generated by the gluon exchange interactions in the quark 
model~\cite{IK}.

%
%%%%%%%%%%%%%%%%%%%%%%%% Theoretical Frame %%%%%%%%%%%%%%%%%%%%%%%%
%
\section{Theoretical Frame}

The chiral constituent quark approach for meson photoproduction 
is based on the low energy QCD Lagrangian~\cite{MANOHAR}
\begin{equation}\label{eq:Lagrangian}
{\cal L}={\bar \psi} \left [ \gamma_{\mu} (i\partial^{\mu}+ V^\mu+\gamma_5
A^\mu)-m\right ] \psi + \dots
\end{equation}
where $\psi$ is the quark field  in the $SU(3)$ symmetry,
$ V^\mu=(\xi^\dagger\partial_\mu\xi+\xi\partial_\mu\xi^\dagger)/2$ 
and 
$A^\mu=i(\xi^\dagger \partial_{\mu} \xi -\xi\partial_{\mu} \xi^\dagger)/2$ 
are the vector and axial currents, respectively, with $\xi=e^{i \Pi f}$; 
$f$ is a decay constant and $\Pi$ the Goldstone boson field.
  
The four components for the photoproduction of
pseudoscalar mesons based on the QCD Lagrangian are:
\begin{eqnarray}\label{eq:Mfi}
{\cal M}_{fi}&=&{\cal M}_{seagull}+{\cal M}_s+{\cal M}_u+{\cal M}_t
\end{eqnarray}
The first term in Eq.~(\ref{eq:Mfi}) is a seagull term. It is generated by the gauge 
transformation of the axial vector $A_{\mu}$ in the QCD Lagrangian.
This term, being proportional to the electric charge of the outgoing mesons, does 
not contribute to the production of the $\eta$-meson.
The second and the third terms correspond to the {\it s-} and {\it u-}channels,
respectively. 
The last term is the {\it t-}channel contribution and contains two parts: 
{\it i)} charged meson exchanges which are proportional to the charge of outgoing 
mesons and thus do not contribute to the process $\gamma N\to \eta N$;  
{\it ii)} $\rho$ and $\omega$ exchange in the $\eta$ production which are 
excluded here due to the duality hypothesis~\cite{LS-2}.

The {\it u-}channel contributions are divided into the nucleon Born
term and the contributions from the excited resonances.  The matrix 
elements for the nucleon Born term are given explicitly, while the 
contributions from the excited resonances above 2 GeV for a given parity 
are assumed to be degenerate so that their contributions could be 
written 
in a compact form~\cite{zpli}.

The contributions from  the {\it s-}channel resonances can be written as
\begin{eqnarray}\label{eq:MR}
{\mathcal M}_{N^*}=\frac {2M_{N^*}}{s-M_{N^*} \big [ M_{N^*}
-i\Gamma(q) \big ]}
e^{-\frac {{k}^2+{q}^2}{6\alpha^2_{ho}}}{\mathcal A}_{N^*},
\end{eqnarray}
where  $k=|\vk|$ and $q=|\vq|$ represent the momenta of the incoming photon 
and the outgoing meson respectively, $\sqrt {s}$ is the total energy of 
the system, $e^{- {({k}^2+{q}^2)}/{6\alpha^2_{ho}}}$ is a form factor 
in the harmonic oscillator basis with the parameter $\alpha^2_{ho}$ 
related to the harmonic oscillator strength in the wave-function, 
and $M_{N^*}$ and $\Gamma(q)$ are the mass and the total width of 
the resonance, respectively.  The amplitudes ${\mathcal A}_{N^*}$ 
are divided into two parts~\cite{zpli,zpl97}: the contribution 
from each resonance below 2 GeV, the transition amplitudes of which 
have been translated into the standard CGLN amplitudes in the harmonic 
oscillator basis, and the contributions from the resonances above 2 GeV
treated as degenerate, since little experimental information is available
on those resonances.

The contributions from each resonance to $\eta$
photoproduction is determined by introducing~\cite{LS-1} a new set of 
parameters $C_{{N^*}}$ and the following substitution rule for the 
amplitudes ${\mathcal A}_{{N^*}}$,
\begin{eqnarray}\label{eq:AR}
{\mathcal A}_{N^*} \to C_{N^*} {\mathcal A}_{N^*} ,
\end{eqnarray}
so that 
\begin{eqnarray}\label{MRexp}{\mathcal M}_{N^*}^{exp} = C^2_{N^*}
 {\mathcal M}_{N^*}^{qm} ,
\end{eqnarray}
where ${\mathcal M}_{N^*}^{exp}$ is the experimental value of 
the observable, and ${\mathcal M}_{N^*}^{qm}$ is calculated in the 
quark model~\cite{zpl97}. 
The $SU(6)\otimes O(3)$ symmetry predicts
$C_{N^*}$~=~0 for ${S_{11}(1650)} $, ${D_{13}(1700)}$, and 
${D_{15}(1675)} $ resonances, and $C_{N^*}$~=~1 for other
resonances in Table~1.  
Thus, the coefficients $C_{{N^*}}$ give a measure of the discrepancies 
between 
the theoretical results and the experimental data and show the extent 
to which the $SU(6)\otimes O(3)$ symmetry is broken in the process 
investigated here.
%
%
%%%%%%%%%%%%%%%%%%%%%%%%%%%%  TABLE I
%   
\begin{table}\label{tab:Res}
\label{assign}
\caption{Resonances included in our study with their 
$SU(6)\otimes O(3)$ configuration assignments.}
\begin{center} 
\begin{tabular}{|ccc|ccc|} \\ 
\hline 
\multicolumn{1}{|c}{Resonance}  & 
\multicolumn{2}{c|}{$~SU(6)\otimes O(3)$}  &
\multicolumn{1}{c}{Resonance}  & 
\multicolumn{2}{c|}{$~SU(6)\otimes O(3)$} \\
%&&&&& \\  
 & State & $C_{N^*}$ & & State & $C_{N^*}$ \\
%&&&&& \\  
\hline        
$S_{11}(1535)$&$N(^2P_M)_{\frac 12^-}$ & 1 & $S_{11}(1650)$&$N(^4P_M)_{\frac 12^-}$ & 0 \\[1ex]
$P_{11}(1440)$&$N(^2S^\prime_S)_{\frac 12^+}$ & 1 &$P_{11}(1710)$&$N(^2S_M)_{\frac 12^+}$ & 1 \\[1ex]
$P_{13}(1720)$&$N(^2D_S)_{\frac 32^+}$ & 1 &$P_{13}(1900)$&$N(^2D_M)_{\frac 32^+}$ & 1 \\[1ex] 
$D_{13}(1520)$&$N(^2P_M)_{\frac 32^-}$ & 1 &$D_{13}(1700)$&$N(^4P_M)_{\frac 32^-}$ & 0 \\[1ex]
$F_{15}(1680)$&$N(^2D_S)_{\frac 52^+}$ & 1 &$F_{15}(2000)$&$N(^2D_M)_{\frac 52^+}$ & 1 \\[1ex]
 &&& $D_{15}(1675)$&$N(^4P_M)_{\frac 52^-}$ & 0 \\ 
\hline     
\end{tabular}
\end{center} 
\end{table}
%
%%%%%%%%%%%%%%%%%%%%%%%%%%%%%%%%%%%%%%%%%%%%%%%%%%%% 

One of the main reasons that the $SU(6)\otimes O(3)$ symmetry is
broken is due to the configuration mixings caused by the one gluon
exchange~\cite{IK}. 
Here, the most relevant configuration mixings are those of the
two $S_{11}$ and the two $D_{13}$ states around 1.5 to 1.7 GeV. The 
configuration mixings can be expressed in terms of the mixing angle
between the two $SU(6)\otimes O(3)$ states $|N(^2P_M)>$  and 
$|N(^4P_M)>$, with the total quark spin 1/2 and 3/2,  
%
%%%
%
\begin{eqnarray}\label{eq:MixS}
\left(\matrix{|S_{11}(1535)> \cr
|S_{11}(1650)>\cr}\right) &=&
\left(\matrix{ \cos \theta _{S} & -\sin \theta _{S}\cr
\sin \theta _{S} & \cos \theta _{S}\cr}\right) 
\left(\matrix{|N(^2P_M)_{{\frac 12}^-}> \cr
|N(^4P_M)_{{\frac 12}^-}>\cr}\right),  
\end{eqnarray}
%
%%%
%
and  
%
%%%
%
\begin{eqnarray}\label{eq:MixD}
\left(\matrix{|D_{13}(1520)> \cr
|D_{13}(1700)>\cr}\right) &=&
\left(\matrix{ \cos \theta _{D} & -\sin \theta _{D}\cr
\sin \theta _{D} & \cos \theta _{D}\cr}\right)
\left(\matrix{|N(^2P_M)_{{\frac 32}^-}> \cr
|N(^4P_M)_{{\frac 32}^-}>\cr}\right).  
\end{eqnarray}
%
%%%
% 

To show how the coefficients $C_{N^*}$ are related to the mixing angles, 
we express the amplitudes ${\mathcal A}_{N^*}$ in terms of the 
product of the meson and  photon transition amplitudes:
\begin{eqnarray}\label{eq:MixAR}
{\mathcal A}_{N^*} \propto <N|H_m| N^*><N^*|H_e|N>,
\end{eqnarray}
where $H_m$ and $H_e$ are the meson and photon transition operators,
respectively. Using Eqs.~(\ref{eq:MixS}) to~(\ref{eq:MixAR}), 
for the resonance ${S_{11}(1535)}$ one finds
\begin{eqnarray}\label{eq:MixAS1}
{\mathcal A}_{S_{11}(1535)} & \propto &  \bigg [ <N|H_m 
|N(^2P_M)_{{\frac 12}^-}> \cos \theta_{S} -
 <N|H_m|N(^4P_M)_{{\frac 12}^-}> \sin \theta_{S} \bigg ] \nn \\
 & &  \bigg [ <N(^2P_M)_{{\frac 12}^-}|H_e|N>  \cos \theta_{S}-
<N(^4P_M)_{{\frac 12}^-}|H_e|N>\sin \theta_{S} \bigg ].
\end{eqnarray}
Due to the Moorhouse selection rule,
the amplitude $<N(^4P_M)_{{\frac 12}^-}|H_e|N>$
vanishes~\cite{zpl97} in our model.
So, Eq.~(\ref{eq:MixAS1}) becomes
\begin{eqnarray}\label{eq:Mix4a}
{\mathcal A}_{S_{11}(1535)}
 & \propto & \bigg [ <N|H_m|N(^2P_M)_{{\frac 12}^-}> 
<N(^2P_M)_{{\frac 12}^-}|H_e|N>\bigg ] \nn \\ 
 & &\bigg [ \cos ^2 \theta_{S} - \sin \theta_{S}
\cos \theta_{S} \frac {<N|H_m|N(^4P_M)_{{\frac 12}^-}>}
{<N|H_m||N(^2P_M)_{{\frac 12}^-}>} \bigg ] .
\end{eqnarray}
%%%%%%%%%%%%%%%%%%%%%%%%%%%%%%
%
where $<N|H_m|N(^2P_M)_{{\frac 12}^-}> <N(^2P_M)_{{\frac 12}^-}|H_e|N>$
determines~\cite{zpl97} the CGLN amplitude for the 
$|N(^2P_M)_{{\frac 12}^-}> $ state, and the ratio
\begin{eqnarray}\label{eq:MixR}
{\cal {R}} =  \frac {<N|H_m|N(^4P_M)_{{\frac 12}^-}>}
{<N|H_m|N(^2P_M)_{{\frac 12}^-}>},
\end{eqnarray} 
is
a constant determined by the $SU(6)\otimes O(3)$ symmetry. Using the 
same meson transition operator $H_m$ from the Lagrangian as in deriving 
the CGLN amplitudes in the quark model, we find ${\cal {R}}$~=~-1 for the $S_{11}$
resonances and $\sqrt{1/10}$ for the $D_{13}$ resonances.
Then, the configuration mixing coefficients can be related to the
configuration mixing angles 
\begin{eqnarray}
C_{S_{11}(1535)} &=& \cos {\theta _{S}} ( \cos{\theta _{S}} - 
\sin{\theta _{S}}),\label{eq:MixS15} \\
C_{S_{11}(1650)} &=& -\sin {\theta _{S}} (\cos{\theta _{S}} + 
\sin{\theta _{S}}),\label{eq:MixS16} \\
C_{D_{13}(1520)} &=& \cos \theta _{D} (\cos\theta _{D} - 
\sqrt {1/10}
\sin\theta _{D}),\label{eq:MixD15} \\
C_{D_{13} (1700)} &=& \sin \theta _{D} (\sqrt {1/10}\cos\theta _{D} + 
 \sin\theta _{D}).\label{eq:MixD17}
\end{eqnarray}
%
%%%%%%%%%%%%%%%%%%%%%%%% Results and Discussion %%%%%%%%%%%%%%%%%%%%%%%%
%
\section{Results and Discussion}
Our effort to investigate the $\gamma p \to \eta p$ process
has gone through three stages.

In our early work~\cite{LS-1}, we took advantage of the differential 
cross section data from
MAMI~\cite{Mainz} 
(100 data points for $E_\gamma ^{lab}$= 0.716 to 0.790 GeV)
and polarization asymmetries measured with polarized target at 
ELSA~\cite{elsa} 
(50 data points for $E_\gamma ^{lab}$= 0.746 to 1.1 GeV)
and polarized beam at GRAAL~\cite{graal}
(56 data points for $E_\gamma ^{lab}$= 0.717 to 1.1 GeV).
Those data allowed us to study the reaction mechanism in the
first resonance region and led to the conclusion~\cite{LS-1}
that the $S_{11}(1535)$ plays by far the major role in this
energy range.

Later, differential cross section data were released by the
GRAAL collaboration~\cite{Graal2000} 
(244 data points for $E_\gamma ^{lab}$= 0.732 to 1.1 GeV). 
Using the four data sets we extended our investigations to the
second resonance region and performed a careful treatment
of the configuration mixing effects. This work~\cite{LS-2} led us to the 
conclusion that the inclusion of a new $S_{11}$ resonance was 
needed to interpret those data.

Finally, the third resonance region has just been covered by the 
CLAS g1a cross section measurements~\cite{CLAS2001}  
(192 data points for $E_\gamma ^{lab}$= 0.775 to 1.925 GeV).

Within our approach, we are in the process of interpreting all available 
experimental results and report here our preliminary findings.

Below, we summarize the main ingradients of the starting point and the 
used procedure leadind to the models ${\cal M}_1$, ${\cal M}_2$,
and ${\cal M}_3$ (see Table~2 and Figs. 1 to 3):
\begin{itemize}
{\item {\bf Mixing angles:} Our earlier works~\cite{LS-1,LS-2} have shown 
the need to go beyond the exact $SU(6)\otimes O(3)$ symmetry. To do
so, we used the relations (12) to (15) for the $S_{11}$ and $D_{13}$
resonances and left the mixing angles $\theta_{S}$ and $\theta_{D}$
as free parameters. 
}
{\item {\bf Model ${\cal M}_1$:} This model includes explictly all 
the eleven known relevant resonances (Table~1) with mass below 2 GeV, 
while the contributions from the known excited resonances above 2 GeV for a 
given parity are assumed to be degenerate and hence written in a 
compact form~\cite{zpli}.
}
{\item  {\bf Model ${\cal M}_2$:} Because of the poor agreement between
the model ${\cal M}_1$ and the data above $E_\gamma ^{lab} \approx$ 1 GeV,
as explained below, and given our previous findings~\cite{LS-2}, 
we introduce a third $S_{11}$ resonance with three free parameters 
(namely the resonance mass, width, and strength).
}
{\item {\bf Model ${\cal M}_3$:} To improve further the agreement between 
our results and the data, we introduce a third  $P_{13}$ 
missing resonance 
with three additional adjustable parameters.
}
{\item {\bf Fitting procedure:} The free parameters of all the above 
three models 
have been extracted using the MINUIT minimization code from the 
CERN Library.
The fitted data base contains roughly 650 values: differential 
cross-sections from MAMI~\cite{Mainz}, GRAAL~\cite{Graal2000},
and JLab~\cite{CLAS2001}, and the polarization asymmetry data from 
ELSA~\cite{elsa} and GRAAL~\cite{graal}.
}
\end{itemize}
%
%%%%%%%%%%%%%%%%%%%%%%  FIG 1 %%%%%%%%%%%%%%%%%%%%%%
%%%%%%%%%%%%%%%%%%%%%%
\begin{figure*}[t!]
\begin{center}
\caption{Differential cross section for the process 
$\gamma p \to \eta p$: angular distribution for 
$E_{\gamma}^{\rm lab}$ = 0.775 GeV to 1.725 GeV.
The curves come from the models ${\cal M}_1$ (dotted), ${\cal M}_2$ (dashed),
and ${\cal M}_3$ (full).
Data are from Refs. [7] and [10].
%~\cite{Mainz}  (empty diamonds), and~\cite{Graal2000} (full circles).
}
\vspace{0.2cm}
\psfig{figure=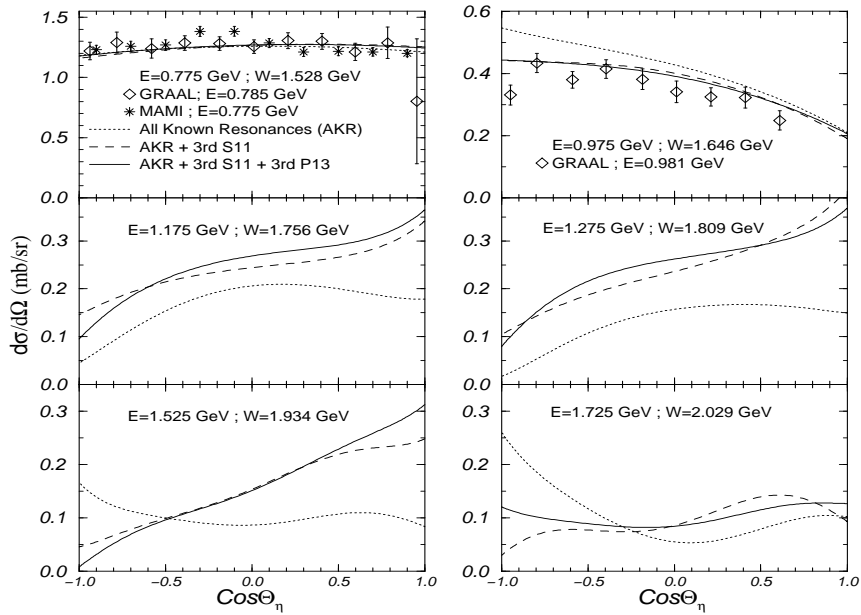,width=12.cm,height=8.3cm}
\vspace{-1.cm}
\label{fig:dsigma1}
\end{center}
\end{figure*}
%%%%%%%%%%%%%%%%%%%%%%%%

In the following, we compare the results of our models with different 
measured observables~\footnote{The differential cross sections from 
JLab~\cite{CLAS2001} were kindly
provided to us by B. Ritchie and M. Dugger and were included in 
our fitted data-base.
However, given that these data have not yet been published by the CLAS
Collaboration, we do not reproduce them here.}.

Figure~1 shows our results at six of the CLAS data energies. 
At the lowest energies we compare our results with data from GRAAL and 
MAMI. As already mentioned, at the lowest energy the reaction mechanism
is dominated by the first $S_{11}$ resonance and the data are equally
well reproduced by the three models. At the next shown energy,
$E_\gamma ^{lab}$=0.975 GeV, we are already in the second resonance 
region and the model ${\cal M}_1$ overestimates the data, while the models
${\cal M}_2$ and ${\cal M}_3$ improve equally the agreement with the 
data.

At $E_\gamma ^{lab}$=1.175 and 1.275 GeV, the model ${\cal M}_1$
underestimates the unshown JLab data (see footnote a). This is also
the case at the two depicted highest energies, except at backward
angles, where again the model ${\cal M}_1$ overestimates the JLab data.
The reduced $\chi^2$ for this latter model is 6.5, see Table~1.

The most dramatic improvement is obtained by introducing a new
$S_{11}$ resonance (Fig.~1, model ${\cal M}_2$), which brings down the 
reduced $\chi^2$ by more than a factor of 2. 

Finally the introduction of a new $P_{13}$ resonance 
(Fig.~1, model ${\cal M}_3$)
gives the best agreement with the data, though it does not play
as crucial a role as the third $S_{11}$ resonance.

Predictions of those models for the total cross section, as well as results
for the fitted polarizations observables are depicted in Figures 2 and 3,
respectively. These theory/data comparisons bolster our conclusions about
the new resonances, without providing further selectivity between models
${\cal M}_2$ and ${\cal M}_3$.
%\vspace{-2cm}
%
%*******
\begin{table}[t!]
%%%%%%%%%%%%%%%%%%%%%%%%%%%%%%%%%%%% 
\begin{center}
\caption{Results of minimizations for the models as explained in the text.}
\vspace{0.2cm}
\begin{tabular}{|ll|rrr|}
%
% & & & &  \\
%
\hline
\multicolumn{2}{|c|}{ parameter}  & 
\multicolumn{1}{c}{${\cal M}_1$}  & 
\multicolumn{1}{c}{${\cal M}_2$}   & 
\multicolumn{1}{c|}{${\cal M}_3$ }   \\
%
% & & & &  \\
\hline
%
% & & & &  \\
%
Mixing angles: & & & &  \\
\multicolumn{1}{|c}{} & \multicolumn{1}{c|}{$\Theta_{S}$} &
\multicolumn{1}{r}{$-37^\circ$} &\multicolumn{1}{r}{$-34^\circ$} &
\multicolumn{1}{r|}{$-34^\circ$}  \\
\multicolumn{1}{|c}{} & \multicolumn{1}{c|}{$\Theta_{D}$} &
\multicolumn{1}{r}{$~~8^\circ$} & \multicolumn{1}{r}{$~11^\circ$} &
\multicolumn{1}{r|}{$~11^\circ$}  \\[1ex]
%
% &  & & &  \\
%Third $ S_{11}$ : &  & & &  \\
Third $ S_{11}$ &  Mass (GeV)  & &1.795 & 1.776 \\
 &   Width (MeV)  & & ~~350 & ~~268 \\[1ex]
%
% &  & & &  \\
%Third $ P_{13}$ &  & & &  \\
Third $ P_{13}$ &   Mass (GeV)  & & & 1.887 \\
 &   Width (MeV)  & & & ~~225 \\[1ex]
%&  & & &  \\
\hline 
 & $\chi^2_{d.o.f}$ & 6.5 & 3.1 & 2.7 \\
\hline
\end{tabular}
\end{center}
\end{table}
%*******

%
%%%%%%%%%%%%%%%%%%%%%%  FIG 2 %%%%%%%%%%%%%%%%%%%%%%
%%%%%%%%%%%%%%%%%%%%%%
\begin{figure*}[t!]
\begin{center}
\caption{Total cross section as a function of total center-of-mass
energy for the process $\gamma p \to \eta p$; curves 
and data as in Fig.~1. 
%~\cite{Mainz}  (empty diamonds), and~\cite{Graal2000} (full circles).
}
\vspace{0.2cm}
\psfig{figure=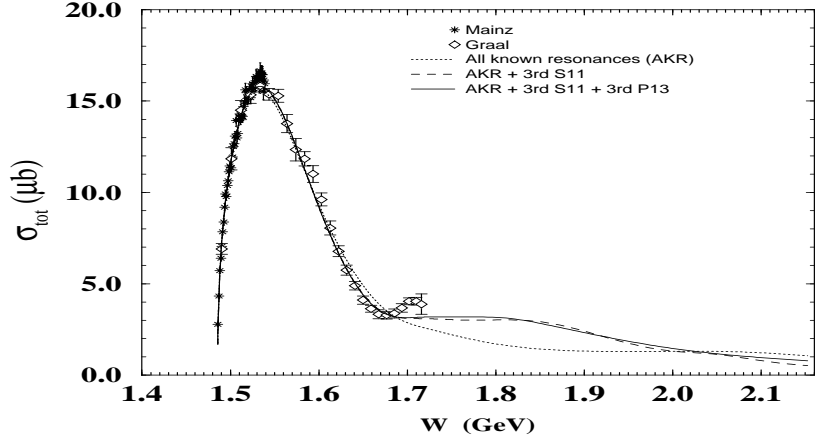,width=12.cm,height=6.5cm}
\vspace{-1.cm}
\label{fig:tot}
\end{center}
\end{figure*}
%%%%%%%%%%%%%%%%%%%%%%%
%
%%%%%%%%%%%%%%%%%%%%%%  FIG 3 %%%%%%%%%%%%%%%%%%%%%%
%%%%%%%%%%%%%%%%%%%%%%
\begin{figure*}[b!]
\begin{center}
\vspace{-.8cm}
\caption{ Single polarization asymmetries angular distribution
for the process 
$\gamma p \to \eta p$; curves as in Fig.~1.
Data are from Refs. [8] and [9].
%~\cite{Mainz}  (empty diamonds), and~\cite{Graal2000} (full circles).
}
\vspace{0.2cm}
\psfig{figure=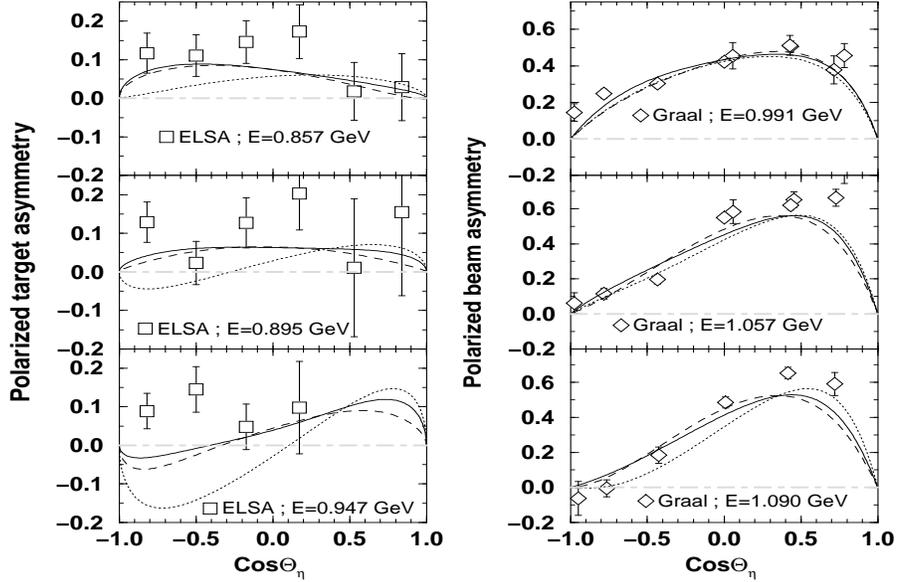,width=12.cm,height=8.cm}
\vspace{-0.6cm}
\label{fig:asym}
\end{center}
\end{figure*}
%%%%%%%%%%%%%%%%%%%%%%%
%%%%%%%%%%%
%

Here, we would like to comment on the values reported in Table 1.

{\bf Mixing angles :} The extracted values for mixing angles 
$\theta _{S}$ and $\theta _{D}$ are identical for models 
${\cal M}_2$ and ${\cal M}_3$ and differ by 3$^\circ$ from those of
the model ${\cal M}_1$. These values are in agreement 
with angles determined by Isgur-Karl model~\cite{IK} and by 
large-$N_c$ approaches~\cite{Nc}.

{\bf Third $S_{11}$ resonance:} 
The extracted values for the mass and width of a new $S_{11}$ are close 
to those predicted by the authors of Ref.~\cite{LW96}
(M=1.712 GeV and $\Gamma_{T}$=184 MeV), 
and our previous findings~\cite{LS-2}.
Moreover, for the one star $S_{11}(2090)$ resonance~\cite{PDG}, the 
Zagreb group
coupled channel analysis~\cite{Zagreb} produces the following values
M = 1.792 $\pm$ 0.023 GeV and $\Gamma_T$ = 360 $\pm$ 49 MeV.
The BES Collaboration
reported~\cite{BES} on the measurements of the
$J/\psi \to p \overline{p} \eta$ decay channel. 
In the latter work, a partial wave analysis
leads to the extraction of the mass and width of the 
$S_{11}(1535)$ and $S_{11}(1650)$ resonances, and the authors find  
indications for an extra resonance with 
$M =$ $1.800 \pm 0.040$ GeV, and $\Gamma_{T} =$ $165^{+165}_{-85}$ MeV.
Finally, a recent work~\cite{GSV} based on the hypercentral constituent 
quark model
predicts a missing $S_{11}$ resonance with M=1.861 GeV.

{\bf Third $P_{13}$ resonance:}
The above mentioned hypercentral CQM predicts also three $P_{13}$ 
resonances with M=1.816, 1.894, and 1.939 GeV.
Finally a relativized pair-creation $^{3}P_0$ approach~\cite{Cap92}
predicts four missing $P_{13}$ resonances in the relevant energy 
region with masses betwenn 1.870 and 2.030 GeV.
%
%%%%%%%%%%%%%%%%%%%%%%%% Results and Discussion %%%%%%%%%%%%%%%%%%%%%%%%
%
\section{Concluding remarks}
We reported here on a study of the process $\gamma p \to \eta p$ for 
$E_{\gamma}^{lab}$
between threshold and $\approx$ 2 GeV, using a chiral constituent quark approach.

We show how the symmetry breaking coefficients $C_{N^*}$ are expressed in terms of 
the configuration mixings in the quark model, thus establish a direct connection
between the photoproduction data and the internal quark gluon structure of baryon 
resonances. The extracted configuration mixing angles for the $S$ and $D$ wave resonances
in the second resonance region using a more complete data-base are in good agreement 
with the Isgur-Karl model~\cite{IK} predictions for the configuration 
mixing angles based on the one gluon exchange, as well as with results
coming from the large-$N_c$ effective field theory based 
approaches~\cite{Nc}.

However, one of the common features in our investigation of $\eta$ photoproduction
at higher energies is that the existing S-wave resonances can not accommodate the
large S-wave component above $E_{\gamma}^{lab} \approx$ 1.0 GeV region. 
Thus, we introduce a third S-wave 
resonance in the second resonance region suggested in the literature~\cite{LW96}. 
The introduction of this new resonance, improves significantly
the quality of our fit and reproduces very well the cross-section increase
in the second resonance region.
The quality of our semi-prediction for the total cross-section and our results
for the polarized target and beam asymmetries, when compared to the data, gives confidence
to the presence of a third $S_{11}$ resonance, for which we extract some static and
dynamical properties: $M \approx$ 1.776 GeV, $\Gamma_{T} \approx$ 268 MeV.
These results are in very good agreement with those in 
Refs. ~\cite{GSV,LW96,BES}, and compatible
with ones in Ref.~\cite{Zagreb}. Finally, we find indications for
a missing $P_{13}$ resonace~\cite{Cap92,GSV} with $M \approx$ 1.887 GeV, $\Gamma_{T} \approx$ 225 MeV.
%
%%%%%%%%%%%%%%%%%%%%%%%% Concluding remarks %%%%%%%%%%%%%%%%%%%%%%%%
%

\section*{Acknowledgments}
One of us (BS) wishes to thank the organizers for their kind invitation 
to this very stimulating symposium.
We are intebted to Barry Ritchie and Michael Dugger for 
having provided us with the CLAS g1a data prior to publication. 

%
%%%%%%%%%%%%%%%%%%%%%%%% Acknowledgments %%%%%%%%%%%%%%%%%%%%%%%%
%

%        \Journal{\}{}{}{}.

\end{document}